\documentclass[dvips]{jltp}
\usepackage[latin1]{inputenc}
\usepackage[dvips]{graphicx}
\usepackage{amssymb}

\title{In--Plane and Out--of--Plane Optical Properties of 
$\mathrm{\mathbf{NdBa_2Cu_3O_{x}}}$ Single Crystals close to x = 7.0}
\author{R.~Hauff$^*$, M.~Göppert$^\star$, B.~Obst$^*$, Th.~Wolf$^*$, P.~Schweiss$^\ddag$,\\
 and H.~Wühl$^{*,\dag}$}
\address{Forschungszentrum Karlsruhe, $^*$ITP and $^\ddag$INFP, 76021 Karlsruhe, Germany\\
Universität Karlsruhe, $^\dag$IEKP and $^\star$AP, 76128 Karlruhe, Germany}

\runninghead{R. Hauff \textit{et al.}}{Optical Properties of $\mathrm{\mathbf{NdBa_2Cu_3O_{x}}}$}
\begin{document}
\maketitle
\begin{abstract}
We present results of reflectivity measurements with $\mathbf{E} \| \mathbf{c}$ and 
$\mathbf{E} \perp \mathbf{c}$ on $NdBa_2Cu_3O_x$ (Nd123) single crystals close to full oxygen doping.
 Along the 
c--axis the optical conductivity $\sigma_1^c$ shows a well developed
 absorption band around 
450 cm$^{-1}$ at all temperatures. 
The in--plane optical properties are dominated by crystal--field excitations at low energies,
 a prominent step at 400 
cm$^{-1}$ and a weaker feature between 500 - 550 cm$^{-1}$. A comparison of $\sigma_1^c$ and the in--plane  scattering rate 
$\Gamma_{ab}(\omega)$ with neutron scattering derived spin susceptibility 
spectra suggests that the in--plane and out--of--plane anomalies are 
caused by the same mechanism, probably electron--spin scattering. 

\end{abstract}
Besides its low--frequency suppression, the normal state c--axis conductivity 
$\sigma_1^c(\omega)$ of underdoped 
Y123 shows two other prominent features, namely the development of an absorption band 
around 450 cm$^{-1}$ and an anomalous softening of the oxygen bending phonon at 320 cm$^{-1}$ 
which is accompanied by a loss of spectral weight\cite{homes,PRB52schuetzmann}.
Although we were able to show that 
the absorption band is not related to the suppression of the conductivity\cite{hauff96}, 
its nature is still not clear. So far, a variety of explanations have been suggested, 
e.g.~regarding it as a new phonon\cite{homes}, a bilayer 
plasmon\cite{grueninger} or as being caused by spin--phonon coupling\cite{hauff96}.
On the other hand, in Y123 the in--plane conductivity $\sigma_1^{ab}$ also shows an anomaly 
in that frequency range\cite{tanner}, and there may be a connection 
between these in--plane and out--of--plane anomalies.

We performed reflectivity measurements both parallel and perpendicular to the 
c--axis of Nd123 single crystals. The Nd123 system was chosen because good quality 
crystals are available and it is possible to study the effects caused 
by the substitution of the bigger Nd ion for the Y ion.

The $\mathrm{NdBa_2Cu_3O_x}$ samples were grown as described elsewhere\cite{prb56wolf}
 and showed the usual twinning. The oxygen content x was adjusted 
by annealing in flowing oxygen at different pressures and temperatures. This yielded x 
= 6.99 and x = 6.96, determined by isobars after Lindemer
 et al.\cite{lindemer95} and cross--checked by neutron scattering. The specific--heat $\mathrm{T_c}$
was 95.5 K (93.5 K) for the
 x = 6.99 (6.96) crystal.
EDX analysis found that 8 \% of the Nd sites were occupied by Y, probably 
caused by the crucible. As a result our
samples showed a slightly higher $\mathrm{T_c}$ than Y--free Nd123 crystals with the same oxygen 
content\cite{erb}. However, apart from this $\mathrm{T_c}$ effect, we regard the Y contamination 
as not crucial 
for the present study.
Due to its small dimension in c--direction the $\mathrm{NdBa_2Cu_3O_{6.99}}$ crystal was 
measured only 
in the ab--plane without using a polarizer. The spectra of the 
$\mathrm{NdBa_2Cu_3O_{6.96}}$ sample were taken on an (a,b)c--plane with both $\mathbf{E}\|\mathbf{c}$ and 
$\mathbf{E}\perp\mathbf{c}$. The measurements were performed in a 
Fourier transform spectrometer in the frequency range 40 - 7000 cm$^{-1}$ and at 
10 - 300 K.  The spectral region of 3000 - 40000 cm$^{-1}$ was 
measured in a prism spectrometer at room temperature. As reference a gold and an aluminum 
mirror have been used.

\begin{figure}[t!]
\includegraphics[width=5in]{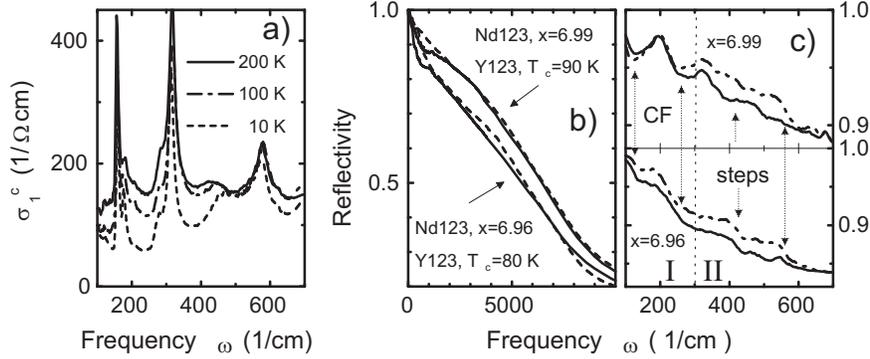}
\caption{a) c--axis conductivity $\sigma_1^c$ of $\mathrm{NdBa_2Cu_3O_{6.96}}$.
 b) in--plane reflectivity of
 $\mathrm{NdBa_2Cu_3O_x}$ with x = 6.99 and x = 6.96 (solid lines) at T = 100 K compared to the
  results for Y123 with 
 $\mathrm{T_c}$´s of 90 K and 
 80 K (dashed lines, Ref.\protect\onlinecite{tanner}). c) in--plane reflectivity 
 of Nd123 at 100 K (solid lines) and 10 K (dashed lines). Region I is 
 dominated by crystal--field (CF) excitations, in region II there are steps around 400 
 cm$^{-1}$ and 500 - 550 cm$^{-1}$.}
\label{Fig1}
\end{figure}

Fig.\ref{Fig1}a shows the Kramers--Kronig calculated c--axis optical conductivity 
of $\mathrm{NdBa_2Cu_3O_{6.96}}$ at various temperatures. The 
main features of the spectra are a strong absorption band around 450 
cm$^{-1}$ far above $\mathrm{T_c}$ and the lack of normal--state anomalies of the oxygen bending 
phonon at 320 cm$^{-1}$. This 
last observation clearly speaks against an interpretation of the absorption
 band as a new phonon 
growing on the expense of other modes as suggested by Homes et 
al.\cite{homes}. Furthermore, it shows that in contrast to what we claimed 
earlier\cite{hauff96}, the absorption band is probably not related to the phonon system at 
all. 
On the other hand, the fact that both the absorption 
band and the phonon anomalies simultaneously disappear in Zn--substituted 
crystals\cite{hauff96} indicates 
that, although the two phenomena are not causally related to each other, they still might 
originate from the same mechanism, e.g.~by coupling to the same kind of excitations. 
Finally, we want to mention that the above results are different to what can be found in 
$\mathrm{YBa_2Cu_3O_x}$ compounds: firstly, in the latter the absorption band develops for temperatures 
$\lesssim$ 150 K, whereas in Nd123 it can be observed up to room 
temperature (not shown). Secondly, in Y123 this band is well known in underdoped samples 
with x $\lesssim$ 6.7, in 
samples with x = 6.96, however, it has never been observed. This indicates that despite the 
high oxygen content the hole 
doping level of the $\mathrm{NdBa_2Cu_3O_{6.96}}$ sample is relatively low which also shows 
in the in--plane reflectivity (Fig.\ref{Fig1}b and c). In Fig.\ref{Fig1}b we compare
 the reflectivity of 
both Nd123 samples at 100 K with the results of Orenstein et al.\cite{tanner} on 
twinned
Y123. In the mid infrared the fully oxygenated 
Nd123 sample behaves like a Y123 
crystal with $\mathrm{T_c}$ = 90 K, whereas the $\mathrm{NdBa_2Cu_3O_{6.96}}$ crystal resembles Y123 with a $\mathrm{T_c}$ of only 
80 K, 
albeit with a less developed shoulder around 4000 cm$^{-1}$. As this band is usually ascribed to 
excitations in the CuO--chains\cite{tanner}, this may indicate that in oxygen reduced Nd123
 the chains are 
more fragmented than in Y123 resulting in a lower hole concentration which is 
in accord to Widder et al.\cite{widder95}. In the far infrared (FIR)
the reflectivity of both Nd123 crystals shows several 
characteristic features (Fig.\ref{Fig1}c). The FIR can be divided
into two regions: region I is dominated by features at 160 
 and 300 cm$^{-1}$, in region II there are steps in the reflectivity around 400 
 and 550 cm$^{-1}$ which are visible at all temperatures becoming more pronounced 
at lower temperatures. We ascribe the region--I features to Nd$^{3+}$ crystal--field 
excitations as they appear almost exactly at the same positions as calculated by Martin et 
al.\cite{martin99}. Concerning region II, similar structures 
have been observed in a variety of compounds and have been 
explained in terms of a coupling of the mid--infrared continuum to 
longitudinal optical (LO) c--axis phonons\cite{PRL69reedyk}. However, this 
interpretation can not apply to our spectra, because the structures are 
also visible in the reflectivity measured on an (a,b)c--surface, a geometry
 where a coupling to LO c--axis modes is forbidden.

\begin{figure}[b!]
\begin{center}
\includegraphics[height=72mm]{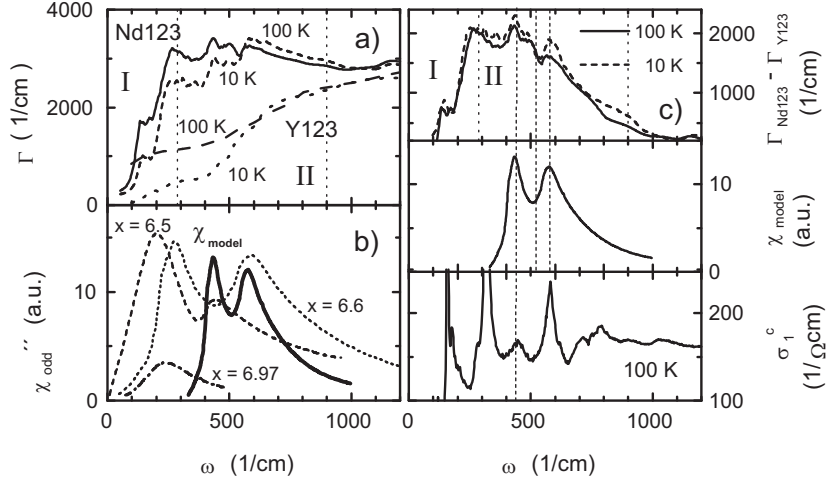}
\end{center}
\caption{a) In--plane scattering rate $\Gamma_{ab}$ of $\mathrm{NdBa_2Cu_3O_{6.96}}$ and fully doped Y123 
(Ref.\protect\onlinecite{liu99}). b) Low temperature spin susceptibility $\chi$" of 
$\mathrm{YBa_2Cu_3O_x}$ at various x (Ref.\protect\onlinecite{bourgeshayden}). The right peak develops only 
at low T. $\chi_{model}$ (thick 
line) is a model 
susceptibility we assume for $\mathrm{NdBa_2Cu_3O_{6.96}}$. c) Difference of $\Gamma_{ab}$ 
for Nd123 and Y123 of a) compared to $\chi_{model}$ and $\sigma_1^c(100 K)$.}
\label{Fig2}
\end{figure}

As an interpretation for this region--II features we rather suggest scattering of the 
electrons 
by spin excitations as described below. Using an 
one--component analysis we calculated for the $\mathrm{NdBa_2Cu_3O_{6.96}}$ sample 
the low--temperature frequency--dependent in--plane scattering rate 
$\Gamma_{ab}$ shown in Fig.\ref{Fig2}a. Additionally, we plotted $\Gamma_{ab}$ of twinned Y123 
($\mathrm{T_c}$ = 92 K) after Liu et al.\cite{liu99} which shows the in these samples generally 
observed non--linear suppression below 1000 cm$^{-1}$. In fully doped untwinned 
Y123 this suppression is only seen in the b--axis component of $\Gamma$\cite{wang98},
 in underdoped material there is also a suppression
  in the a--axis component\cite{puchkov}. Comparing the Nd123 and Y123 spectra of 
Fig.\ref{Fig2}a suggests that this behaviour is also present in 
Nd123, but masked between 100 - 1000 cm$^{-1}$ by strong additional contributions. 
 In order to extract these contributions, we used 
 $\Gamma_{ab}$(Y123) as a background and subtracted it from $\Gamma_{ab}$(Nd123) yielding the
  difference spectrum of Fig.\ref{Fig2}c. Hereby, we assumed that 
  $\Gamma_{ab}$(Y123) describes the generic decrease, free of spin scattering 
  contributions which seems to be reasonable considering the weak spin spectrum of fully doped 
  Y123 (Fig.~\ref{Fig2}b).
 In region I the enhancement of $\Gamma_{ab}$ is probably caused by the 
 crystal--field excitations mentioned above, which have 
 nothing to do with the following scenario and are therefore not further considered, while 
 we ascribe the  
 enhanced scattering rate in region II to electron--spin interaction.
  We assume a model 
 spin susceptibility 
 $\chi_{model}$ for $\mathrm{NdBa_2Cu_3O_{6.96}}$ (Fig.\ref{Fig2}b and c) which possesses the same
 double--peak structure as the 
 neutron--scattering derived spectra $\chi$" of underdoped Y123\cite{bourgeshayden}, but 
 has a gap at low energies which accounts for the fact, that with 
 increasing oxygen content the low--energy spin excitations disappear\cite{bourgeshayden}. 
 The size of the gap may be justified if we consider
  that the big Nd ion causes an expansion of the unit cell compared to Y123 which
   may alter the spin exchange 
  interaction in the Nd system resulting in a larger gap.
  With this $\chi_{model}$ we are able to reproduce the two 
 peaks at 400 and 550 cm$^{-1}$ in the difference spectrum of Fig.\ref{Fig2}c as well as 
 the peak in the 
 conductivity $\sigma_1^c$. In this scenario the left peak in $\chi_{model}$, 
 present already at 300 K, causes the absorption band, whereas the 320 
 cm$^{-1}$ phonon lying in the gap region is not affected.
 In Y123, on the other hand, the left peak of the measured $\chi$" roughly coincides with the 320 
 cm$^{-1}$ phonon and the right peak 
 with the absorption band position. This then may lead to the observed phonon anomalies and the 
 growth of the absorption band when the 
 left peak shifts to lower energies and the right peak develops with decreasing 
 temperature as seen by Bourges et al.\cite{bourgeshayden}.

In summary, by optical measurements we found anomalies in the in--plane and out--of--plane 
properties of highly oxygenated Nd123. The anomalies are similar to the ones observed in 
underdoped Y123 and are probably caused by electron--spin interaction.

\end{document}